\documentclass[twocolumn,pra,aps,amssymb]{revtex4}
\usepackage{physics}
\usepackage{graphics,epstopdf}
\usepackage{graphicx}
\graphicspath{ {images/} }
\usepackage[colorlinks=true, citecolor=blue, urlcolor=blue ]{hyperref}

\usepackage{algorithm2e}
\usepackage{amsmath,amssymb,amsfonts,amsthm}
\usepackage{multirow}
\usepackage{verbatim}
\usepackage{url}
\usepackage{comment}
\usepackage{mathtools}

\usepackage{epsf,pgf,graphicx}
\usepackage{color}
\usepackage{tikz,pgf}

\begin{document}
\title{Revisiting Integer Factorization using Closed Timelike Curves\footnotetext{Part of this work was done while the first two authors were visiting R. C. Bose Centre for Cryptology and Security, Indian Statistical Institute, Kolkata during the Summer of 2017 (between the 6th and the 7th semesters of their Bachelor of Engineering course in Electronics \& Telecommunication Engineering Department of Jadavpur University) for internship under the supervision of the third author.}
}


\author{Soumik Ghosh$^1$, Arnab Adhikary$^2$ and Goutam Paul$^3$}
\affiliation{$^1$Institute for Quantum Computing,\\
                University of Waterloo, ON N2L 3G1, Canada.\\
                Email: soumik.ghosh@uwaterloo.ca\\
$^2$Centre for High Energy Physics,\\
                Department of Physics,\\
                Indian Institute of Science, Bangalore 560 012, India.\\
              Email: arnaba@iisc.ac.in\\
$^3$Cryptology and Security Research Unit,\\
R. C. Bose Centre for Cryptology and Security,\\
Indian Statistical Institute, Kolkata 700 108, India.\\
Email: goutam.paul@isical.ac.in\\
}

\begin{abstract}
Closed Timelike Curves are relativistically valid objects allowing time travel to the past. Treating them as computational objects opens the door to a wide range of results which cannot be achieved using non relativistic quantum mechanics. Recently, research in classical and quantum computation has focused on effectively harnessing the power of these curves. In particular, Brun (Found. Phys. Lett., 2003) has shown that CTCs can be utilized to efficiently solve problems like factoring and QSAT (Quantified Satisfiability Problem). In this paper, we find a flaw in Brun's algorithm and propose a modified algorithm to circumvent the flaw.
\end{abstract}
\maketitle
\date{}

\section{Introduction}
Closed Timelike Curves (CTCs) arise as one of the admissible solutions \cite{Godel, Bonnor, Gott} to Einstein’s field equations of the general theory of relativity. Although not proven to exist, they opened the doors for possible time travel to the past and assorted paradoxes. Ever since, physicists \cite{Morris, Hartle, Throne} have pondered over their ramifications and causal violations. Their existence was thought to be highly implausible as they seemed to violate the chronology protection conjecture \cite{Hawking}. The situation changed when David Deutsch, in his revolutionary paper \cite{Deutsch}, imposed a self-consistency condition for CTCs, resolving the causal violations and giving scientists an information theoretic tool to study such curves. His model for CTCs became popularly known as Deutschian Closed Timelike Curves (D-CTCs). Later, Bennett and Schumacher \cite{Schumacher} suggested an alternative nonequivalent formulation of CTCs using quantum teleportation and post-selection (P-CTCs). This was further developed by Seth Lloyd \cite{Lloyd2} and experimentally simulated \cite{Lloyd} by Aephraim Steinberg's group.
 
 In recent years, a new branch of research has sprung up to fathom the implications of such curves, modelling them with tools from information theory. They have been shown to possess enormous computing power and can lead to counter-intuitive solutions to complex computational problems. They are particularly interesting as they allow non-linear quantum mechanics \cite{Gisin, Abrams, Cassidy}, which has interesting ramifications. 
 
 Dave Bacon \cite{Bacon} showed that a quantum computer with access to D-CTCs can solve NP-complete problems using only polynomial number of computations. Scott Aaronson \cite{Aaronson} showed that if D-CTCs are part of reality, quantum computers are no more computationally efficient than classical Turing machines. 

Brun \cite{Brun1} showed a quantum computer equipped with a D-CTC can detect non orthogonal states. There were also been efforts to clone quantum states \cite{Ahn} in the presence of CTCs. DeJonghe et al.~\cite{DeJonghe} proved the evolution of the chronology respecting part of a D-CTC can be a discontinuous function of the initial state. Pati et al.~\cite{Pati} showed it was not possible to purify mixed states of qubits traversing a D-CTC while still being consistent. The results of Brun et al.~\cite{Brun1} imply D-CTCs can break the Holveo \cite{Holevo} bound and violate the uncertainty principle. Bennett et al.~\cite{Bennett1} has questioned some of these striking results, suggesting the circuits of \cite{Aaronson, Brun1} do not work \cite{Bennett2} when acting on a classically labeled mixture of states. 

More recent papers by \cite{Ralph} have studied CTCs using the Heisenberg picture. It has also been shown that a density matrix formulation \cite{Cavalcanti} is not valid for a nonlinear theory. P-CTCs have also been extensively studied, their advantage being the fact that they are consistent with path integral \cite{Lloyd} formulations. By invoking Aaronsons's result \cite{Aaronson2}, it has been shown \cite{Lloyd} that the computational power of P-CTCs is equivalent to that of the complexity class PP. Aaronson's recent work \cite{Aaronson4} has aimed at understanding what is computable by a CTC. 

Additionally, recent works \cite{Ringbauer, Yuan, WildeBrun} have aimed at simulating closed timelike curves and replicating their results. There has also been significant research on flow of information \cite{Dunlap, Indranil}, causality considerations \cite{Ujjwal, Wallman, LoboP, Korotaev, Baumler}, no-go theorems \cite{Sami}, state preparation \cite{Pati2}, and thermodynamic considerations \cite{Bartkiewicz} within CTCs. CTCs and entanglement \cite{Moulick, Jung, Gedik, Pati3} is an equally interesting avenue for research.   
 
\section{CTCs and factorization}
 Before the flurry of works in Closed Timelike Curves, Todd Brun \cite{Brun2} developed an algorithm that would allow us to find out factors of a number in constant time using CTCs. In the same paper, he extended the general framework of his algorithm to solve NP-Hard problems like Quantified Satisfiability Problem (QSAT). The same framework was later used by him in his study of P-CTCs \cite{Brun3}. In this paper, we study the framework and mainly build on Brun's work. 

An important remark here is that Brun \cite{Brun2} never explicitly mentioned which type of Closed Timelike Curve he used in his computations.This is a significant issue in his paper \cite{Brun2} which is also mentioned by Aaronson and Watrous \cite{Aaronson}. However, we have noted the prevalence of model-agnostic approaches like Brun's in our later sections.

Brun uses paradoxes to construct his arguments as to why the algorithms work. It can't be a Deutschian model \cite{Deutsch} as a straightforward grandfather paradox is not enough to cause a contradiction in the Deutschian model. Brun's model~\cite{Brun2} for CTCs only requires self-consistency to be preserved at all costs which is more in line with P-CTCs.
 
\subsection{Brun's Factoring Algorithm} 
An example of using CTCs to factor numbers was proposed by Todd Brun in \cite{Brun2}, given in Algorithm~\ref{algof1}.

\begin{algorithm}[htbp]
%\SetLine
\KwIn{Any number $N$}
\KwOut{A factor of $N$}
\BlankLine
\nl~~ $input(N)$\;     
\nl~~ $timeRegister \gets -1$\;
\nl~~ $t \gets clock( )$\;
\nl~~ $p \gets timeRegister$\;

\nl~~ \If {$(p > 1)$ and $(N \equiv 0 \mod{p})$}
{
\nl~~ $goto$ $FINAL$\;
}
\nl~~ $JUMP:$ $p \gets 1$\;
\Repeat { $(N \equiv 0 \mod{p})$ or $(p > \sqrt{N})$ }
{
\nl~~ $p \gets p + 1$\;
}
\nl~~ \If{$(p > \sqrt{N})$}
{
\nl~~ $p \gets N$\;
}
\nl~~~~ $FINAL: timeSet(t,p)$\;
\nl~~~~ $output(p)$\;
\nl~~~~ $end$\;
\caption{Factoring algorithm as proposed by \cite{Brun2}}
\label{algof1}
\end{algorithm}

\subsection{Working of the algorithm}
The algorithm has a register $timeRegister$ whose value can be set from the future to a particular integer. Initially, the variable is initialized to $1$. The time instance is stored in $t$. Initially, as the condition of line $6$ is not satisfied, the algorithm proceeds from line $8$ to line $14$.

By now, a factor of the input is stored in $p$. If the input is a prime number, $p$ stores the number itself. To generalize, $p$ stores the result of factorization.

\subsubsection{Use of a CTC}
After storing the result in $p$, the CTC is used to alter the value of the variable $timeRegister$ from the future. In line $15$, $timeSet(t, p)$ changes the value of the $timeRegister$ at time instance $t$ to $p$. This has interesting ramifications.

Since, at time instance $t$, the value of $timeRegister$ contains the result, it now satisfies the condition of line $5$, as $N$ is now divisible by $p$.  
Thus, the algorithm branches to line $15$. Thereafter, it displays the output $p$.

The algorithm doesn't actually traverse the  $repeat$ loop even once. By clever use of a CTC and by threatening to produce a contradiction if the $repeat$ loop is traversed, the algorithm has the result stored in $p$ from time instance $t$ itself. Without a CTC, one needs to iterate through the entire computational block, as is done usually, before arriving at the result.

The same strategy can be utilized for other difficult computations.  By going back in time and storing the result of the computation in a special variable, we can ensure that the actual computational block is avoided. This solution, though non-intuitive, is perfectly self-consistent.

\subsection{The flaw in the algorithm}
The main claim of the author is that a consistent solution to the problem will mean that the variable $timeRegister$ will contain the factor of the number, before the execution of the loop. This in turn means that the inner loop won't be executed, giving us the factor as soon as the program is run. This is the only self-consistent situation although quite counter-intuitive. 

However, there can be another consistent solution. The variable $timeRegister$, instead of storing a non-trivial factor of the number, can store instead the number itself. Hence, there is no surety that the algorithm will give us the result we desire once it is executed. 

The algorithm can work if we modify the $if$ condition in line $5$ to reject the cases where $p$ is equal to $N$. But then, the algorithm will not work for prime numbers. For primes, in absence of other factors, $p$ has to store $N$. 

Basically, the fact that a non-trivial prime factor is found in constant time by just executing the algorithm means that we have effectively erased the process of finding that prime factor, making the output variable free not to take in the desired value of a prime factor.

\section{Our Approach}
While formulating an algorithm using Closed Timelike Curves, care must be taken to avoid spurious solutions which can compromise the working of the algorithm. In the previous section, we have pointed out one such spurious solution in Brun's formulation of the factoring algorithm. Hereafter, we have modified it in a way such that the problem is resolved.

 First, we have explained a flaw in the original algorithm. Then, we have put forward a better solution but with more time complexity. After proposing our solution, we have optimized the solution by introducing one more temporal variable. We have taken a model-agnostic approach whose relevance is discussed in the subsection below.

\subsection{Importance of our model-agnostic approach}
It is true that the works of Brun \cite{Brun2} have been superseded by that of Aaronson and Watrous \cite{Aaronson} who rigorously studied closed timelike curves in the light of complexity theory. However, there are a few differences between their work and ours. 

Primarily, they have based their work on D-CTCs whereas, as stated earlier, the time machine Brun used here is more in line with a model agnostic approach. Such an approach is present in many other works like~\cite{LoboP, Akl} to name a few. Interestingly, \cite{Akl} came after \cite{Aaronson} published his results. 

Evidently, even without assuming a model, one can find interesting qualitative results with regard to complexity and universality in computation by incorporating elements of time travel involved, as has been done by \cite{Moravec, Akl}. This is why Brun’s paper \cite{Brun2} might be relevant even today and why the flaw in his original algorithm needs to be rectified and future researchers warned of the common traps and mistakes.

Additionally, while Aaronson's work contends that closed timelike curves equip both quantum and classical computers with the power of solving PSPACE problems, we are concerned with a particular instance of the complexity class PSPACE, namely, a factoring algorithm. Even after acknowledging the power of classical computers to utilize closed timelike curves in solving problems belonging to PSPACE, and the well-known class NP contained in PSPACE, it is still interesting to understand how it does so, which is achieved by studying the factoring algorithm, which is in NP, in detail. Our approach can be utilized in formulating effective algorithms using CTCs for similar problems in NP, without error.

Moreover, from Aaronson's perspective \cite{Aaronson2, Aaronson3}, his research and that of others, detailing the astonishing prowess of a computer equipped with D-CTCs, can be used as evidence against the existence of CTCs that follow Deutsch's prescription. This has resulted in many nonequivalent formulations of CTCs, from P-CTCs \cite{Lloyd, Lloyd2} to recent works on other models~\cite{TCTC}. For a general approach to study the flow of causality in these curves, a model agnostic approach can yet be relevant.

\section{A modified but inefficient algorithm}
We can modify the previous algorithm as shown, with the help of one more conditional branch. The modified algorithm is presented as Algorithm~\ref{algof2}.

\begin{algorithm}[htbp]
                        %\SetLine
                        \KwIn{Any number $N$}
                        \KwOut{A non trivial factor of $N$}
                        \BlankLine
                        \nl~~ $input(N)$\;     
                        \nl~~ $timeRegister \gets -1$\;
                        \nl~~ $t \gets clock( )$\;
                        \nl~~ $p \gets timeRegister$\;
                        \nl~~ \If{$(p == N)$}
                        {
\nl~~ $goto$ $JUMP$\;
}
\nl~~ \If {$(p > 1)$ and $(N \equiv 0 \mod{p})$}
{
\nl~~ $goto$ $FINAL$\;
}
\nl~~ $JUMP:$ $p \gets 1$\;
\Repeat { $(N \equiv 0 \mod{p})$ or $(p > \sqrt{N})$ }
{
\nl~~ $p \gets p + 1$\;
}
\nl~~ \If{$(p > \sqrt{N})$}
{
\nl~~ $p \gets N$\;
}
\nl~~ $FINAL: timeSet(t,p)$\;
\nl~~ $output(p)$\;
\nl~~ $end$\;
                        \caption{Modified factoring algorithm}
                        \label{algof2}
\end{algorithm}

\subsection{Discussion on correctness}
In this algorithm, we keep an additional check to determine whether $p=N$. If, it is true, then, we force the algorithm to execute the computational block from line $11$. If the number is composite, the block spits out a factor of the number which goes back in time via $timeRegister$ to modify the variable $p$, making $p$ contain that factor from our chosen instance. 

However, if $p$ did contain the factor in the first place, the condition $p=N$ wouldn't have been valid. This gives us a contradiction. By threatening to produce this contradiction, we always ensure that the consistent solution for composite numbers is never the number itself but instead, a non-trivial factor of that number.

\subsection{Problem with prime numbers}
As we saw in the last section, for composite numbers, the main computational block would not run. The algorithm would take constant time $O(1)$ to be implemented. 

For prime numbers, we force the algorithm to execute the main computational block. However, the computational block from line $11$ gives out the number itself. We go back in time and store the number in $timeRegister$. This produces no contradictions anywhere in algorithm and $p=N$, in the case of prime numbers, is indeed the consistent result. 

But, as a trade-off, for every prime number, the computational block will run with a time complexity of $O(\sqrt{N})$. This is because, for a prime number, the block would iterate through lines $12$ to $14$, checking every number till the square root of the input integer, and then go back in time to set $p = N$. Hence, the $if$ condition is satisfied for line $5$ and there is no way to avoid the $repeat$ loop even with access to a CTC. The $repeat$ loop has worst case complexity $O(\sqrt{N})$, which becomes the worst case complexity for the entire program.

In case of the original algorithm, the time complexity, for both prime and composite numbers, was $O(1)$. What we achieve in terms of correctness is countervailed by a lack of efficiency.

\section{An optimal algorithm using CTC}
We can design a better algorithm with O(1) complexity with a little computational trickery and the use of an extra register whose value can be modified from the future. It is shown in Algorithm~\ref{algof3}.

\begin{algorithm}[htbp]
                        %\SetLine
                        \KwIn{Any number $N$}
                        \KwOut{A non trivial factor of $N$}
                        \BlankLine

\nl~~ $input(N)$\;
\nl~~ $timeRegister \gets -1$\;
\nl~~ $flagRegister \gets 0$\;
\nl~~ $t \gets clock( )$\;
\nl~~ $p \gets timeRegister$\;
\nl~~ $f \gets flagRegister$\;
\If {$(p == N)$ and $(f==0) $}
{
\nl~~ $goto$ $JUMP$\;
}
\If {$(p > 1)$ and $(N \equiv 0 \mod{p})$}
{
\nl~~ $goto$ $FINAL$\;
}
\nl~~ $JUMP:$ $p \gets 1$\;
\Repeat {$(N \equiv 0 \mod{p})$ or $(p > \sqrt{N})$}
{
\nl~~ $p \gets p + 1$\;
}
\If{$(p > \sqrt{N})$}
{
\nl~~ $p \gets N$\;
\nl~~ $flagSet (t,1)$\;
}
\nl~~ $FINAL: timeSet(t,p)$\;
\If {($p == N$)}
{
\nl~~ $flagSet (t,1)$\;
}
\nl~~ $output(p)$\;
\ $end$\;
                        \caption{Optimal factoring algorithm}
\label{algof3}
                \end{algorithm}

\subsection{Working of the Algorithm}
The challenge is to provide an algorithm which works optimally for both prime and composite numbers. For this, we use two registers which can be accessed from the future, a $timeRegister$ and a $flagRegister$. In this case, if $N$ is composite but $p$ stores $N$ itself, we threaten to produce the same contradiction as in the previous section.

However, if the number is prime, we force the program to enter the main computational block, where we change the value of the $flagRegister$ from $0$ to $1$. The condition of entering the block in the first place was the flag value being $0$. Hence, once we change the value of $timeRegister$ and $flagRegister$, a contradiction is produced unless the algorithm avoids the main computational block for prime numbers, as it did in Brun's original algorithm. This reduces the time complexity for prime numbers to $O(1)$.

For composite numbers, the consistent solution will be $f=0$ and $p= factor$, while for prime numbers, the consistent solution will be $f=1$ and $p=N$.

\section{Conclusion}
In this paper, we point out a flaw in Brun's original factoring algorithm using CTCs~\cite{Brun2} and propose a remedy. The modified final algorithm can now easily be used as a recursive tool in factoring large numbers or in factorization in the presence of CTCs with finite length, as Brun has demonstrated in his paper~\cite{Brun2}.

The case study of the factoring algorithm gives us a caveat to be kept in mind in the future while writing computer codes utilizing self-consistent time travel to the past, as there is always a chance of encountering extraneous fixed points which disturb the actual solution we want to achieve. However, the extraneous fixed points, as have been shown, can be eliminated, at least for the factorization problem, using some more insight. It remains to be seen whether the same holds good for other tasks where a similar issue is encountered. Further research in this domain can focus on how and whether quantum algorithms can speed up by CTCs. Development of a Universal Turing Machine with the help of CTCs could also be an interesting future work.

\end{document}